# Long-range orbital transport and inverse orbital Hall effect in Co/Ru-based terahertz emitters

Chao Zhou [1†], Shaohua Zhang[1†], Lei Hao[1†], Yaxuan Jin[1], Xianguo Jiang[1], Ning Yang[1], Li Zheng[1], Hao Meng[1], Chao Lu[1], Wendeng Huang[1], Yizheng Wu[2,3*], Yan Zhou[4*], Jia Xu[1*]

[1.] Department of Physics, School of Physics and Telecommunication Engineering, Shaanxi University of Technology, Hanzhong 723001, China

[2.] Department of Physics and State Key Laboratory of Surface Physics, Fudan University, Shanghai 200433, China

[3.] Shanghai Research Center for Quantum Sciences, Shanghai 201315, China

[4.] Guangdong Basic Research Center of Excellence for Aggregate Science, School of Science and Engineering, The Chinese University of Hong Kong, Shenzhen, Shenzhen, Guangdong 518172, China

[†]These authors contributed equally to this work.

## Abstract

The utilization of terahertz (THz) emission spectroscopy in femtosecond photoexcited spintronic heterostructures has emerged as a versatile tool for investigating ultrafast spin-transport in a noncontact and non-invasive manner. However, the investigation of ultrafast orbital-transport is still in the primitive stage. Here, we experimentally demonstrate the orbital-to-charge current conversion in Co/Ru heterostructures. Time-domain measurements reveal delayed and broadened terahertz waveforms with increasing Ru thickness, consistent with long-range orbital transport. In Co/Pt/Ru trilayers, the terahertz emission is further enhanced through constructive interference between the inverse spin Hall effect (ISHE) in Pt and inverse orbital Hall effect (IOHE) in Ru, while reversed stack structures show suppressed output. Ferromagnetic resonance (FMR) measurements reveal a strong correlation between damping and THz amplitude, highlighting efficient angular momentum conversion. These results position Co/Ru as a promising orbitronic platform for tunable ultrafast THz emission. Our results not only strengthen the physical mechanism of condensed matter physics but also pave the way for designing promising spin-orbitronic devices and terahertz emitters.

# I. Introduction

The conversion of spin currents into charge currents has become a foundational mechanism in spintronics, underpinning a wide range of ultrafast and energy-efficient devices [1-5]. Among these, spin-orbit coupling (SOC)-based spin-to-charge conversion (SCC), particularly the inverse spin Hall effect (ISHE), has been extensively utilized in terahertz (THz) spintronic emitters and magnetic sensing [6-9]. In such systems, heavy metals with strong SOC, such as Pt, W, and Ta serve as the conversion layers, where laser-excited spin currents from an adjacent ferromagnet are converted into transverse charge currents that emit THz radiation [2, 10, 11]. However, the efficiency of ISHE-based THz emitters is limited by the short spin diffusion length of only a few nanometers in most heavy metals [11-14]. As a result, ultrathin heavy metal layers are required to achieve efficient SCC while simultaneously reducing THz absorption. At the same time, the overall emission efficiency is also affected by the Fabry-Pérot interference effects [15]. The entangled interplay among these factors has posed significant challenges for the optimization of THz sources with high efficiency.

More recently, an alternative pathway known as orbital-to-charge conversion (OCC) has emerged, offering new degrees of freedom, material flexibility, and the capability to transport on much longer distances [6, 16-19]. This mechanism, encompassing the inverse orbital Hall effect (IOHE) and inverse orbital Rashba-Edelstein effect (IOREE), enables the conversion of orbital angular momentum currents into charge currents [19-22]. OCC-based processes are particularly attractive because they do not rely on strong SOC materials. Unlike spin Hall conductivity, the orbital Hall conductivity (OHC) in many 3d and 4d transition metals is orders of magnitude larger and supports longer diffusion lengths [6, 16, 18]. This opens opportunities for THz emitters and spin-orbit devices with broader material choices and new functionalities.

To effectively generate IOHE-induced THz emission, two key material conditions must be satisfied. On one hand, the ferromagnetic (FM) layer must exhibit strong spin-orbital conversion coefficient ($\eta_{L-S}$), as found in elements such as Ni, Co, and their alloys (CoPt, CoFe) [17, 21, 23]. These materials efficiently convert spin currents to orbital angular momentum during laser excitation, and can be used as a strong source of orbital currents. On the other hand, the adjacent nonmagnetic (NM) metal must possess a large orbital Hall conductivity, enabling efficient conversion of orbital currents into charge currents. Several light metals, including Ti, Cu, Zr, and so on, have recently been identified as promising candidates due to their large OHC despite weak SOC [18, 19].

Ruthenium (Ru), a 4d transition metal, presents an intriguing yet underexplored platform in this context. Recent work has shown that introducing Ru layers atop magnetic insulators like YIG can significantly enhance magnetic damping, implying strong orbital angular momentum absorption [24]. Additionally, Ru has been observed to influence spin transport properties in metallic heterostructures, and generate angular momentum accumulation that can manipulate FM spins [24-26]. Despite these findings, very few studies have investigated Ru-based structures for THz emission [5, 11], but its potential as an orbital current sink has not been experimentally validated.

In this work, we systematically investigate Co/Ru heterostructures to explore their efficiency as orbitronic THz emitters. Leveraging the strong spin-orbital interconversion of Co and the expected large orbital Hall conductivity of Ru, we demonstrate that Co/Ru bilayers produce robust THz emission predominantly via IOHE. Through systematic thickness-dependent measurements and direct comparison with Co/Pt bilayers, we confirmed the long-distance transport of orbital transport in Ru. We further uncover a cooperative enhancement in THz signals when Pt is introduced between Co and Ru, allowing ISHE and IOHE to constructively enhance the THz output. The impact of stacking order is also examined, revealing that the spatial alignment and directionality of spin and orbital currents critically influence emission efficiency. Ferromagnetic resonance (FMR) measurements further confirmed the importance of film stacking order in improving the efficiency of OCC. Our results establish Co/Ru as a viable and tunable platform for orbital-driven ultrafast spin-orbit photonics, and offer new opportunities for designing efficient spin-orbit optoelectronic devices.

## II. Experiments

Bilayer and trilayer samples consisting of Co, Ru, and Pt layers were fabricated on 1-mm-thick glass substrates by magnetron sputtering at room temperature. For deposition, DC sputtering was used for the Pt layer, while RF sputtering was employed for the Co and Ru layers. The deposition rates were approximately 2 nm/min for Co and Ru, and 1 nm/min for Pt. The base pressure in the sputtering chamber was $5.0 \times 10^{-7}$ Torr. The layer thickness was controlled by deposition time and varied between 0 and 50 nm. Sample stacking sequences are described from bottom to top, with the first layer deposited directly on the substrate and the final layer forming the top surface. The numbers in the parenthesis represent the thickness of each layer in nanometer.

THz emission spectroscopy measurements were performed at room temperature in a dry air environment with relative humidity below 5%. A schematic diagram of the THz emission measurement setup is presented in Figures 1a-b. During measurements, samples were mounted in an electromagnet providing a 1000 Oe in-plane magnetic field along the laboratory y-axis. Femtosecond laser pulses were incident normally from the back side, exciting the magnetic heterostructures after passing through the substrates, and the emitted THz signals were detected through time-resolved electro-optic sampling. The THz wave emission arises from the photoexcitation by linearly polarized femtosecond laser pulses from a Ti:Sapphire laser oscillator (with a duration of ~80 fs, a center wavelength of 800 nm, and a repetition rate of 80 MHz). The pulse energy of the laser beam was kept below ~1.7 nJ to ensure that the emitted THz signal was linearly proportional to the laser intensity. The excitation beam diameter was approximately 100 μm. The emitted THz signal was detected using the electro-optic sampling technique with a 0.5 mm-thick electro-optic ZnTe (110) crystal.

For FMR measurements, the back side of the sample was wrapped in aluminum foil, and both the foil and sample were fixed onto the coplanar waveguide (CPW) using adhesive tape. The edge of the sample was along the central signal line of CPW, which was itself oriented parallel to the in-plane external magnetic field. A non-magnetic spacer was used to gently press the sample against the CPW, ensuring stable positioning and efficient microwave coupling without damaging the film surface. A commercial vector network analyzer (VNA) was employed to measure the microwave transmission parameter $S_{21}$. The FMR spectra were extracted from the magnitude of the complex $S_{21}$ parameter, following background subtraction and normalization. This approach allows for clear identification of the resonance features associated with magnetization precession and damping.

## III. Results and discussion

### A. Strong THz Emission and Long-Range Orbital Transport in Co/Ru

To establish a baseline and identify the underlying mechanisms of THz emission in Co/Ru heterostructures, we first compare them with the well-known spintronic emitter consisting of Co/Pt bilayer. In the conventional Co/Pt system (Figure 1a), femtosecond laser excitation induces ultrafast demagnetization in the Co layer, generating a spin current that flows into the adjacent Pt layer. There, due to the large spin Hall angle in Pt, the spin current is converted into a transverse charge current via the ISHE and then emitting a single-cycle THz pulse [11-14].

The Co/Ru bilayer structure (Figure 1b) offers a fundamentally distinct mechanism. While Ru possesses a relatively weak spin Hall effect, theoretical and experimental studies have demonstrated that it hosts a significant OHC [24-29]. Thus, spin–orbit coupling in Co facilitates the conversion of a portion of the ultrafast spin current into orbital angular momentum, allowing both spin and orbital currents to be injected into the Ru layer. Within Ru, the orbital currents can subsequently be converted to charge currents via the IOHE, generating THz radiation even when the ISHE contribution from Ru is negligible.

To test this hypothesis, we measured the THz emission from a Co(1.5)/Ru(30) bilayer, where the Ru thickness far exceeds the typical spin diffusion length, effectively suppressing any contribution from the ISHE [12, 14]. As shown in Figure 1c, a clear THz signal is still observed even at this large Ru thickness, suggesting that the underlying charge conversion mechanism might involve long-range transport and extends over a much longer length scale than typical spin-mediated processes.

To exclude the possibility that the observed signal originates from ultrafast demagnetization in Co, we examined the THz signal emitted from a 1.5-nm Co single-layer film, as shown in Figure 1c. It exhibits only a weak THz signal that does not reverse polarity when pumped from either the front or back side of samples, consistent with ultrafast demagnetization that is not affected by the light incidence direction [18, 23]. In contrast, the Co/Ru bilayer produces an obvious THz signal that fully reverses polarity with either the magnetic field direction or the laser incidence is reversed. Such clear sign-reversal behavior excludes non-directional demagnetization processes and further supports the dominance of orbital-mediated transport and IOHE-based conversion in the Co/Ru system. In addition, we also fabricated the inverted Ru/Co structure and performed measurements on both Co/Ru and Ru/Co samples under varying magnetic-field strength and field-rotation angle (See Fig. S1 of the Supplementary Information [30]). The results consistently confirm that the THz emission arises from the conversion of angular-momentum flow into charge current [18, 19].

To further confirm the orbital origin of the THz emission in Co/Ru, we investigate a series of bilayer samples with the form Co(1.5)/Ru($d_{\text{Ru}}$), where the Ru thickness $d_{\text{Ru}}$ varies from 0 to 50 nm. The time-domain THz waveforms are shown in Figure 2a, and the corresponding frequency spectra are included in Figure 2c. Remarkably, the Co/Ru samples maintain a substantial THz signal even when the Ru thickness reaches 50 nm, indicating long-range angular momentum transport well beyond typical spin diffusion lengths. Moreover, we examined the effect of pump-polarization rotation on the THz emission from the Co/Ru bilayer. The THz amplitude ($\Delta V$) exhibits a very weak sinusoidal modulation as the pump polarization angle is varied, indicating the presence of a minor nonlinear optical contribution [31, 32]. However, quantitative analysis reveals that this nonlinear-optical component accounts for only about 7% of the maximum THz amplitude (See Fig. S2 of Supplementary Information [30]). Therefore, the dominant mechanism responsible for THz generation in the Co/Ru bilayer is the conversion of angular-momentum flow into charge current, rather than nonlinear optical rectification.

For comparison, we performed an analogous experiment on Co(1.5)/Pt($d_{\text{Pt}}$) bilayers, varying Pt thickness from 0 to 15 nm. As shown in Figure 2b, the THz signal from Co/Pt rapidly diminishes with increasing Pt thickness and becomes nearly undetectable at $d_{\text{Pt}} \approx 13$ nm, consistent with the expected spin diffusion-limited behavior in conventional ISHE-based emitters [12, 14]. Figure 2c shows the frequency spectra of the time-domain signals in Figures 2a-b, where the signals are mainly distributed in the frequency range of 0~4 THz for all samples.

The thickness dependence of the peak-to-peak THz amplitude ($\Delta V$) for both Co/Pt and Co/Ru bilayers are summarized in Figure 2d, where the definition of $\Delta V$ is marked in Figure 2a. For both Co/Ru and Co/Pt bilayers, the THz signal initially increases with Ru or Pt thickness, peaking around 3~4 nm. However, beyond this peak, the Pt-based signal decays sharply whereas the Ru-based signal declines much more gradually, implying a significantly longer effective propagation length in the Co/Ru system.

Notably, the time-domain waveforms in Figures 2a-b also reveal a key difference in pulse dynamics: the THz peak position remains nearly unchanged for Co/Pt samples but shifts progressively with Ru thickness in Co/Ru bilayers, as marked by dash lines. This behavior is quantified in Figure 2e, where the THz time delay $\tau_D$ is extracted via Hilbert transformation [20], and is plotted as a function of nonmagnetic metal thickness (See details in Fig. S3 of Supplementary Information[30]). While $\tau_D$ is largely independent of Pt thickness, it increases steadily with $d_{\text{Ru}}$ until a critical thickness. The increasing trend of $\tau_D$ vs. Ru thickness $d_{\text{Ru}}$ can be described by the following formula [18]:

$$\tau_D = \frac{\int_{d_{\text{Ru}}/v_o}^{\infty} \exp\left(-\frac{t}{\tau_{\text{of}}}\right) dt}{\int_{d_{\text{Ru}}/v_o}^{\infty} \frac{1}{t} \exp\left(-\frac{t}{\tau_{\text{of}}}\right) dt} \tag{1}$$

where $\tau_{\text{of}}$ is the orbital flip time, $v_o$ is the group velocity of the orbital carriers,

$d_\text{Ru}$ is the thickness of the Ru layer. By fitting this behavior, we extract an effective orbital velocity $v_0$=0.12±0.03 nm/fs and an associated diffusion time $\tau_\text{of}$ in the range of 100 to 200 fs. The orbital diffusion length can then be estimated approximately as the point where the value of $\tau_D$ reaches maximum and starts to decrease [18], with the value of $l_\text{of} \approx 20$ nm which is much larger than the spin diffusion length in Pt [12]. The derived orbital transport parameters of Ru are in good agreement with values reported from spin-torque FMR studies [24, 27, 33] and match recent reports of orbital propagation velocities in similar metallic systems [18, 34].

Finally, Figure 2f presents the extracted THz pulse width as a function of nonmagnetic layer thickness, with the pulse width is defined as indicated in Figure 2b. In Co/Pt, the pulse width remains constant while it grows monotonically with $d_\text{Ru}$ in Co/Ru bilayer, which further confirms that the emitted THz waveform evolves due to orbital diffusion dynamics. These findings collectively demonstrate that the dominant charge conversion mechanism in Co/Ru is governed by the IOHE, distinct from spin Hall processes and consistent with long-distance orbital transport within the Ru layer [6, 16].

**B. Collaborative THz Emission in Co/Pt/Ru Trilayers**

Having established that the THz emission in Co/Ru originates predominantly from the IOHE contribution, we further investigated whether orbital and spin conversion mechanisms can cooperate to enhance THz emission. To this end, a Pt layer with high-efficient SOC was inserted between Co and Ru. As illustrated in Figure 3a, femtosecond laser pulses excite a spin current $j_\text{S}$ in the Co layer, which is injected into the Pt layer. Given the large spin Hall angle of Pt [35], a portion of this spin current $j_\text{S}$ is converted into a transverse charge current $j_\text{C}$ via the ISHE to generate THz radiation. Simultaneously, due to strong SOC of Pt, part of the spin current $j_\text{S}$ is converted into orbital angular momentum current $j_\text{L}$ which is then injected into the Ru layer. There, the IOHE further converts $j_\text{L}$ into a transverse charge current $j_\text{C}$. Since both Pt and Ru have positive spin and orbital Hall angles, respectively, the charge currents $j_\text{C}$ generated via ISHE and IOHE are aligned in the same direction. These two THz contributions therefore interfere constructively and are expected to enhance the total THz emission [19, 36].

This cooperative enhancement is experimentally confirmed in Figure 3b, the Co(3)/Pt(1)/Ru(1) exhibits stronger THz emission than either Co(3)/Pt(1) or Co(3)/Ru(1) bilayers. Notably, as shown in the Figure 3d, the extracted $\Delta V$ of trilayer exceeds the sum of the corresponding Co/Pt and Co/Ru bilayer signals, directly confirming a cooperative contribution from both mechanisms, instead of simple additive effect. However, when the Pt thickness increases to 2 nm, the THz signal from Co(3)/Pt(2)/Ru(1) decreases below that of Co(3)/Pt(2), as shown in Figures 3c-d. This suppression observed at $d_{Pt}$=2 nm can be attributed to enhanced orbital damping within the thicker Pt layer and increased overall trilayer thickness, which leads to signal attenuation due to optical absorption and electrical screening [17, 19].

To further assess the role of the Ru layer, we systematically varied its thickness in the Co(3)/Pt(1)/Ru($d_{Ru}$) configuration. As shown in Figure 3e, the THz amplitude $\Delta V$ initially increases with Ru thickness, peaking around 1 nm. This enhancement corresponds to the cooperative regime, where IOHE in Ru is boosted by spin-to-orbital conversion in Pt layer. Beyond this point, the signal gradually decreases, indicating the increasing influence of orbital damping, optical absorption, and electrical shunting effects. Notably, even at $d_{Ru}$ around 50 nm, the THz signal remains detectable, which highlights the long-range transport nature of orbital currents. The orange-shaded region (I) in Figure 3e marks the regime where cooperative IOHE and ISHE contributions dominate.

To further clarify the impact of layer order and signal symmetry, we examined two other trilayer configurations: Co/Ru/Pt and Ru/Co/Pt, as shown in Figure 4a-b. In the Co(3)/Ru(1)/Pt(1) structure, the THz signal is significantly lower than that of Co(3)/Pt(1) and even similar to Co(3)/Ru(1), as shown in Figure 4c. While the Ru(1)/Co(3)/Pt(1) structure, where the flow of $j_L$ and $j_S$ are different, yields a THz signal slightly lower than Co(3)/Pt(1), it is still stronger than Co(3)/Ru(1) bilayer, as shown in Figure 4d.

We further studied the THz signal amplitude as a function of Ru thickness in both structures, as shown in Figure 4e. For Ru/Co/Pt, the THz signal exhibits a gradual decay and vanishes around $d_{Ru} \approx 30$ nm. However, Co/Ru/Pt shows a rapid decrease with increasing Ru thickness, with the signal disappearing at $d_{Ru}$ of approximately 5 nm. This disparity suggests that the suppression of long-range angular momentum transport in the Co/Ru/Pt structure, is primarily due to interfacial and bulk scattering within the Ru layer. Specifically, in Co/Ru/Pt trilayer, strong spin-orbit scattering at the Co/Ru interface combined with spin current attenuation across the thick Ru layer, severely limits spin transmission to the Pt layer, as shown in Figure 4a. On the other hand, the orbital current from Co is relatively weak to generate strong signals in Ru, and the IOHE in Pt is also weak to generate a strong charge current. One might argue that the spin-to-orbital conversion in Pt should still enable long-distance transport of $j_L$, but in this structure, even if Pt can convert $j_S$ into $j_L$, the direction of flow for $j_L$, is opposite to that for $j_L$ from Co, resulting in even smaller net signals. Additionally, the top Ru and Pt layers decreases the light absorption in Co and increase the overall conductance of the film, further reducing the observed THz signal and resulting in the rapid decay of emission with increasing Ru thickness [19, 36].

While the Ru/Co/Pt configuration enables spatial separation of spin and orbital currents, as shown in Figure 4b. Upon femtosecond excitation, spin current flows upward into Pt and contributes to THz emission via ISHE, while orbital current flows downward into Ru and emits via IOHE [19]. The resulting charge currents are oppositely directed, they partially cancel each other out, leading to reduced overall THz output. Despite this, the orbital current in Ru/Co/Pt also exhibits long-range diffusion, as evidenced by detectable THz emission up to ~30 nm Ru thickness. To quantify the impact of stacking order on angular momentum transmission, we quantitatively compare the decay of the THz emission as a function of Ru thickness for the three trilayer configurations: Co/Pt/Ru, Ru/Co/Pt, and Co/Ru/Pt. Here, we define the attenuation length ($d_h$) as the Ru thickness at which the THz signal amplitude drops to half of its peak value [14]. A visual indication of how $d_h$ is determined is shown in Figure 4e, and the extracted values are summarized in Figure 4f.

Among the three geometries, the Co/Pt/Ru structure exhibits the largest $d_h$ of approximately 12 nm, indicating that orbital currents can propagate efficiently across the Ru layer and contribute to THz generation via IOHE. The Ru/Co/Pt trilayer shows a slightly shorter $d_h$ of ~8 nm, but still maintains substantial orbital current transport capability. In sharp contrast, the Co/Ru/Pt configuration reveals a markedly suppressed $d_h$ of only ~0.4 nm, suggesting that spin currents are rapidly attenuated within the Ru layer before reaching the Pt interface. This pronounced difference highlights the crucial role of interface quality, layer sequence, and the relative efficiency of ISHE and IOHE in mediating long-range angular momentum transfer and THz emission in these heterostructures. These results demonstrate that among the multilayer configurations tested, Co/Pt/Ru provides the most effective structural platform for maximizing the cooperative enhancement between IOHE and ISHE.

**C. FMR Analysis of Damping and Spin Transparency**

To further investigate the spin dissipation and angular momentum transport in Co/Ru-based structures, we performed ferromagnetic resonance (FMR) measurements on several representative bilayer and trilayer samples (See details in Fig. S9 of the Supplementary Information [30]). Figure 5a displays the FMR spectra of Co(3)/Ru(1) bilayer measured at different microwave frequencies. The linewidth broadens with increasing frequency, enabling extraction of the Gilbert damping parameter $\alpha$ [37-39]. The resonance magnetic field $H_R$ as a function of microwave frequency $f$ can be described using the Kittel formula:

$$f = \frac{\gamma}{2\pi}\sqrt{H_R(H_R + M_{eff})} \qquad (2)$$

The $f$-$H_R$ relationships of all samples in the inset of Figure 5b are well overlapped, indicating consistent magnetic properties like saturation magnetization and anisotropy across samples. In Figure 5b, we plot the full width at half maximum (FWHM) of the resonance peaks $\Delta H$ as a function of frequency for various samples studied the previous section.

All samples exhibit linear trends, which can be fitted using the formula:

$$\Delta H = \Delta H_0 + \frac{2\pi\alpha}{\gamma\mu_0}f \tag{3}$$

where $\Delta H_0$ is the inhomogeneous broadening related to the film quality, $\alpha$ is the Gilbert damping coefficient, $\gamma$ is the gyromagnetic ratio, and $\mu_0$ is the permeability of vacuum and the extracted damping parameters are summarized in Figure 5c.

We observe that the damping values correlate well with the corresponding THz signal amplitudes, when the layer thickness of each layer is the same, samples with larger damping (such as Co/Pt and Co/Pt/Ru) also exhibit stronger THz emission. This suggests that these samples possess more efficient angular momentum dissipation channels, either via ISHE or IOHE, which contribute to enhanced ultrafast THz radiation [40]. To quantify the dissipation dynamics in Co/Ru, we further measured the damping $\alpha$ as a function of Ru thickness in Co(1.5)/Ru($d_{Ru}$) bilayers. As shown in Figure 5d, $\alpha$ increases monotonically with $d_{Ru}$. Based on the theory in ferromagnetic /nonmagnetic metal bilayers [41-43], the enhancement of Gilbert damping in a ferromagnet due to a neighboring nonmagnetic layer can be described as:

$$\Delta\alpha = \alpha - \alpha_0 = \frac{g\mu_B g^{\uparrow\downarrow}}{4\pi M_S} \frac{1}{t_{Co}} \left(1 - e^{-\frac{2t_{Ru}}{\lambda_{Ru}}}\right) \tag{4}$$

Fitting the data with Eq. (4) yields the maximum of damping enhancement as $\Delta\alpha = 0.0080 \pm 0.0013$, the damping of Co without Ru is $\alpha_0 = 0.0587 \pm 0.0002$, and an effective orbital diffusion length in Ru of $\lambda_{Ru} = 46 \pm 13$ nm. These values are comparable to prior reports on orbital pumping [24] and confirm that Ru acts as an efficient angular momentum sink in Co/Ru heterostructures, consistent with its strong IOHE-mediated THz emission behavior. The value of $\lambda_{Ru}$ acquired from FMR measurements is much larger than that from the THz measurements, possibly indicating that the orbital current diffusion exhibits different behaviors in the THz and GHz frequencies [44].

## IV. Conclusion

In summary, we demonstrate that Co/Ru heterostructures serve as an efficient platform for orbitronic terahertz emission, where the dominant contribution arises from the IOHE in Ru layer. The THz signal persists even in ultrathick Ru layers and displays spectral broadening and delay evolution that are incompatible with conventional spin Hall mechanisms. We further establish a mechanism for cooperative spin-orbit THz generation by inserting a Pt spacer between Co and Ru. In the optimized Co/Pt/Ru trilayer, the ISHE in Pt and IOHE in Ru produce collinear charge currents that constructively interfere, leading to a clear enhancement in THz output. In contrast, reversed or misaligned structures such as Co/Ru/Pt and Ru/Co/Pt exhibit either spin current attenuation or destructive interference, suppressing the emission. FMR measurements across these systems reveal a consistent correlation between damping and THz amplitude, reinforcing the interpretation that efficient angular momentum dissipation of both spin or orbital, underpins strong ultrafast charge conversion. Our results provide direct evidence for orbital-driven THz emission in transition-metal bilayers and highlight structural engineering as a viable route to tailor and enhance orbitronic functionalities in spintronic THz sources.

**Acknowledgements:**

This work was supported by the National Natural Science Foundation of China (Grant No. 12204295 and 12204296), the Open Research Project of State Key Laboratory of Surface Physics from Fudan University (Grant No. KF2022_17), the Youth Science and Technology Star Program of Shaanxi Province (2025ZC-KJXX-15), the Young Talent Fund of Association for Science and Technology in Shaanxi (20250512), the Youth Project in Sanqin Talent Introduction Program, and the Shaanxi Province Outstanding Young Talent Support Program for Universities and Young Hanjiang Scholar of Shaanxi University of Technology, and Shaanxi University of Technology (Grants No. SLGRCQD044 and SLGRCQD046). The Innovation Fund of Shaanxi University of Technology (Grant No. SLGYCX2514 and SLGYCX2511). Yizheng Wu acknowledges the support from the National Key Research and Development Program of China (Grant No. 2024YFA1408501), the National Natural Science Foundation of China (Grant No.11974079, 12274083, 1222100), the Shanghai Municipal Science and Technology Major Project (Grant No. 2019SHZDZX01), and the Shanghai Municipal Science and Technology Basic Research Project (Grant No. 22JC1400200). Hao Meng acknowledges the support from the National Natural Science Foundation of China (Grant No.12174238). Y. Z. acknowledges support by the Shenzhen Fundamental Research Fund (Grant No. JCYJ20210324120213037), and the National Natural Science Foundation of China (12374123), Guangdong Basic Research Center of Excellence for Aggregate Science, and the 2023 SZSTI stable support scheme. Li Zheng acknowledges the support from the Natural Science Foundation of Shaanxi Provincial Department of Education (24JK0365), Shaanxi University of Technology (SLGRC202419).

**Figures:**

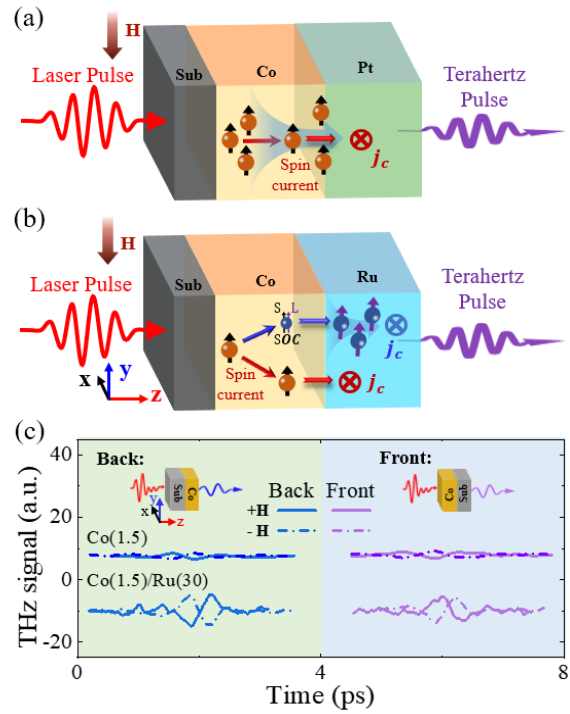

**Figure 1** Sketch illustration of THz emitter based on (a) Co/Pt and (b) Co/Ru heterostructures. (c) Terahertz emission spectra of single-layer Co and Co/Ru bilayer under reversed laser incidence directions and external magnetic field orientations, the inset illustrates the laser incidence geometry with respect to the sample surface.

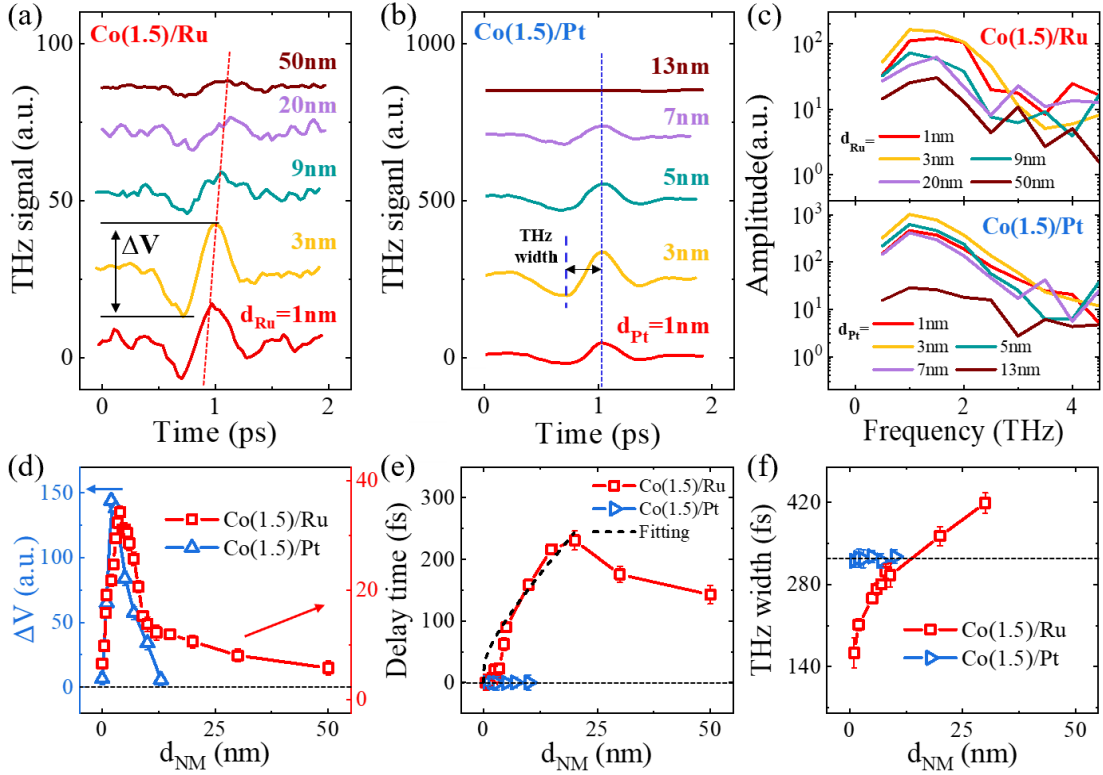

**Figure 2** (a)-(b) Time-domain THz signals as a function of nonmagnetic layers thickness for (a) Co(1.5nm)/Ru and (b) Co(1.5nm)/Pt. (c) Fast-Fourier-transformed (FFT) amplitude spectra corresponding to the time-domain signals in (a) and (b), showing evolution of spectral features with thickness. (d) Dependence of THz signal amplitude ($\Delta V$) on Ru and Pt thicknesses, extracted from (a) and (b). (e) Time delay of THz signals as a function of nonmagnetic layer thickness $d_{\mathrm{Ru}}$ and $d_{\mathrm{Pt}}$, the dashed line represents the best fitting curve with Eq. (1). (f) Evolution of the THz pulse width as a function of Ru and Pt thicknesses, the definition of THz width is illustrated in panel (b).

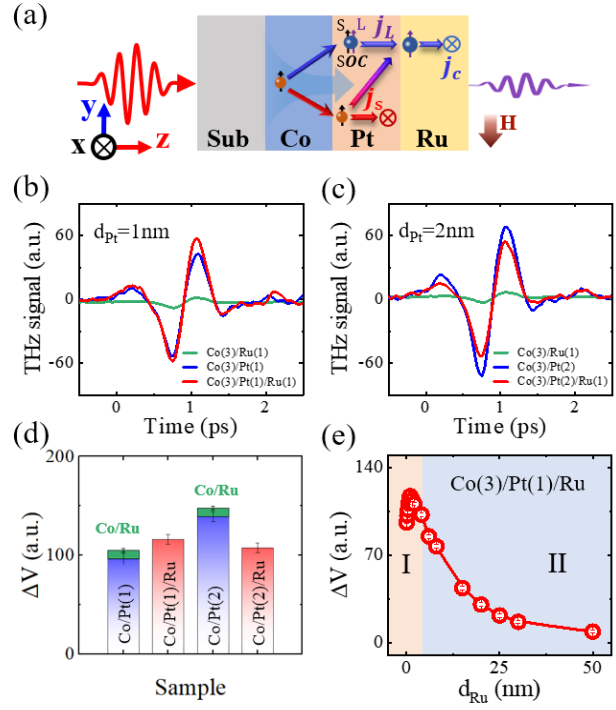

**Figure 3** (a) Schematic of the synergistic THz emission mechanism in Co/Pt/Ru trilayer. Comparison of THz emission for Co(3)/Pt($d_{Pt}$)/Ru(1), Co(3)/Pt($d_{Pt}$), and Co(3)/Ru(1) structures with Pt layer thickness of (b) $d_{Pt}$=1 nm and (c) $d_{Pt}$=2 nm. (d) Extracted THz signal amplitudes corresponding to the samples shown in panels (b) and (c). (e) THz signal amplitude ($\Delta V$) as a function of Ru thickness $d_{Ru}$ for the Co(3)/Pt(1)/Ru($d_{Ru}$) configurations.

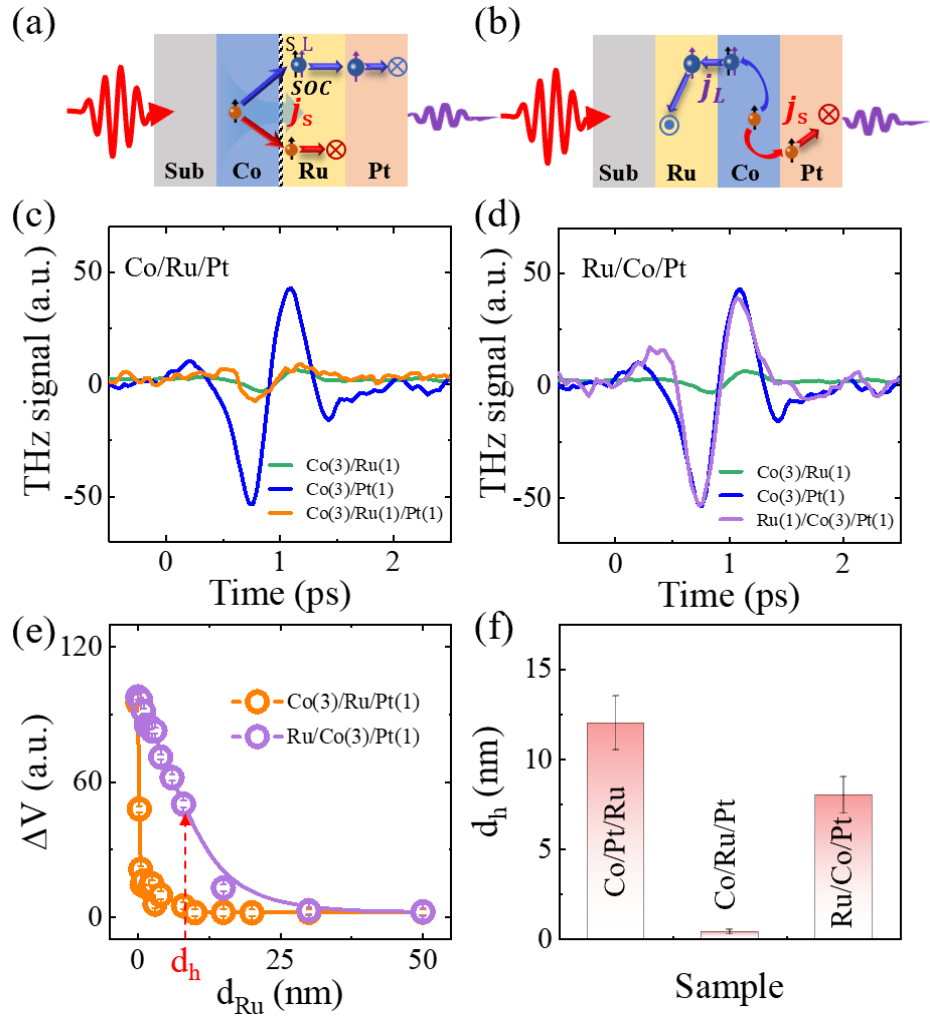

**Figure 4** (a)-(b) Schematic illustration of THz emission mechanism in (a) Co/Ru/Pt (b) Ru/Co/Pt stacks. Time-domain THz signals of (c) Co(3)/Ru(1)/Pt(1), (d) Ru(1)/Co(3)/Pt(1) and corresponding bilayers. (e) THz signal amplitude ($\Delta V$) as a function of Ru thickness $d_{Ru}$ for both trilayers. (f) Comparison of extracted decay lengths $d_h$ for different stacking orders, the $d_h$ is defined in panel (e) at which the THz amplitude $\Delta V$ drops to half its maximum value.

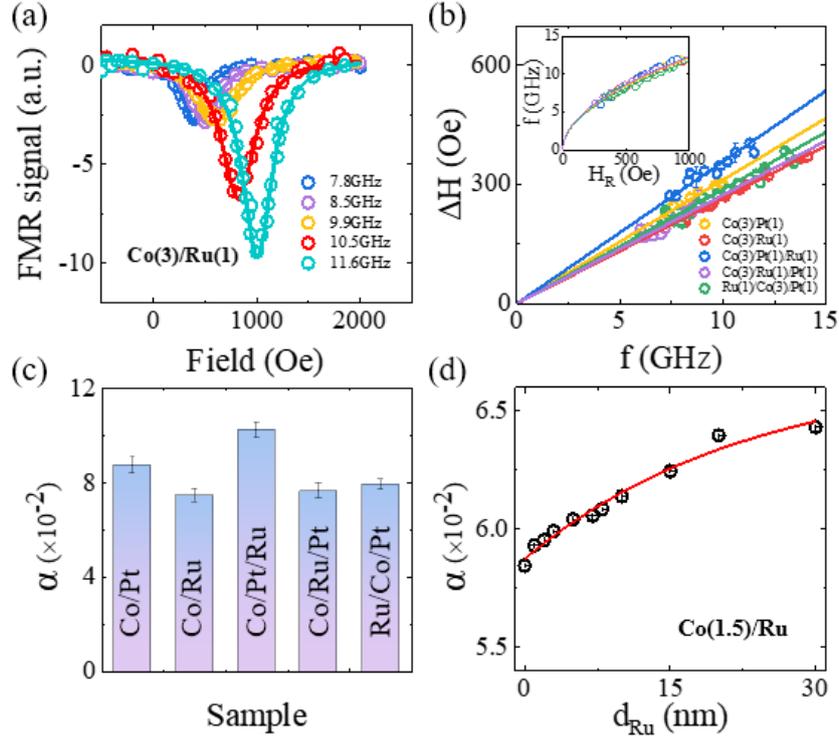

**Figure 5** (a) FMR spectra of Co(3)/Ru(1.5) heterostructure measured at different microwave frequencies. (b) Fitted FWHM $\Delta H$ of the FMR spectra as a function of microwave frequency for samples, the symbols represent the experimental values, and the solid line shows the fit with Eq. (3). The insert shows the dependence of microwave frequency on the resonance magnetic field $H_R$. The symbols show the experimental values and the curves show the best fit with Eq. (2). (c) Comparison of the effective Gilbert damping coefficient ($\alpha$) for Co/Pt, Co/Ru, and Co/Pt/Ru trilayers with different stacking sequences. (d) Evolution of the $\alpha$ as a function of Ru thickness in Co(1.5 nm)/Ru bilayers. The solid curve shows the best fit with Eq. (4).